\documentclass[fleqn,usenatbib]{mnras}

\usepackage{newtxtext,newtxmath}

\usepackage[T1]{fontenc}

\DeclareRobustCommand{\VAN}[3]{#2}
\let\VANthebibliography\thebibliography
\def\thebibliography{\DeclareRobustCommand{\VAN}[3]{##3}\VANthebibliography}

\usepackage{booktabs}  
\usepackage{multirow}
\usepackage{subfig}
\usepackage{graphicx}	
\usepackage{amsmath}	
\usepackage{color}
\usepackage[dvipsnames]{xcolor}

\newcommand{\kms}{km~s\ensuremath{^{-1}}}
\newcommand{\ha}{H$\alpha$}
\newcommand{\hb}{H$\beta$}
\newcommand{\msun}{${\rm M}_{\odot}$}

\newcommand{\mh}{${\rm M}_{\rm H_2}$}
\newcommand{\mhi}{${\rm M}_{\rm HI}$}
\newcommand{\mstar}{${\rm M}_{\ast}$}

\newcommand{\deltamzr}{$\Delta$Z}
\newcommand{\hi}{{H{\sc i}}}

\title[Gas accretion and the mass-metallicity relation]{The impact of gas accretion and AGN feedback on the scatter of the mass-metallicity relation}

\author[N. Yang et al.]{
Nancy Yang\thanks{E-mail: nancy.yang@physics.ox.ac.uk} $^{1,2}$, 
Dirk Scholte $^1$, and Am\'elie Saintonge\thanks{E-mail: a.saintonge@ucl.ac.uk} $^1$\\
$^1$Department of Physics and Astronomy, University College London, London, WC1E 6BT, UK\\
$^2$Sub-department of Astrophysics, Department of Physics, University of Oxford, Denys Wilkinson Building, Keble Road, Oxford, OX1 3RH, UK\\
}

\date{Accepted XXX. Received YYY; in original form ZZZ}

\pubyear{2022}

\begin{document}
\label{firstpage}
\pagerange{\pageref{firstpage}--\pageref{lastpage}}
\maketitle

\begin{abstract}
The gas-phase metallicity of galaxies encodes important information about galaxy evolution processes, in particular star formation, feedback, outflows and gas accretion, the relative importance of which can be extracted from systematic trends in the scatter of the mass-metallicity relation (MZR). Here, we use a sample of low redshift (0.02 < z < 0.055) galaxies from SDSS to investigate the nature of the scatter around the MZR, the observables and physical processes causing it, and its dependence on galaxy mass. We use cold gas masses inferred from optical emission lines using the technique of \citet{scholte23} to confirm that at fixed stellar mass, metallicity and gas mass are anti-correlated, but only for galaxies up to $M_{\ast}= 10^{10.5} M_{\odot}$. In that mass regime, we find a link between the offset of a galaxy from the MZR and halo mass, using the amplitude of the two-point correlation function as a proxy for halo mass; at fixed stellar mass, the most gas-poor galaxies reside in the most massive halos. This observation is consistent with changes in gas accretion rates onto galaxies as a function of halo mass, with environmental effects acting on satellite galaxies also contributing. At higher stellar masses, the scatter of the MZR does no longer correlate with gas or halo mass. Instead, there is some indication of a link with AGN activity, as expected from models and simulations that metallicity is set by the interplay between gas in- and outflows, star formation, and AGN feedback, shaping the MZR and its scatter.

\end{abstract}


\section{Introduction}

The gas-phase metallicity of a galaxy is directly linked to the metal yield of the star formation process and the cycling of gas in and out of the interstellar medium. The metallicity can for example be reduced by accretion of metal-poor gas or the ejection of preferentially metal-enriched gas. Given this fundamental connection with all aspects of the baryon cycle, many of which are challenging to observe directly, the gas-phase metallicity is a powerful tool to constrain galaxy formation and feedback models  \citep[e.g.][]{tinsley80,edmunds90,dave12,ma16}. 

The mass-metallicity relation (MZR) is the scaling between the stellar mass of galaxies and their gas-phase metallicity  \citep[e.g.][see also Fig. \ref{fig:gassmass_comparison}, left]{lequeux79,tremonti04}. The MZR has been characterised up to $z\sim 3$ \citep[e.g.][]{erb06,mannucci09,zahid13,sanders21}, and over 5 orders of magnitude in stellar mass at $z\sim0$ \citep[e.g.][]{lee06,berg12,andrews13,jimmy15,james17}.  The shape of the MZR is determined by the relative balance between metal production, loss and dilution, as a function of galaxy mass and redshift. The shape is characterised by a flattening above a mass of \mstar$\sim 10^{10-10.5}$\msun, with the value of the threshold dependent on metallicity calibration and redshift \citep{kewley08,zahid13}.  

Various models have been invoked to explain the shape of the MZR. Early investigations considered closed-box chemical evolution models \citep{searle72}, or variations thereof that allow for inflow of metal-poor gas and outflow of metal-rich gas \citep[e.g.][]{larson72,larson74}. Several studies identify outflows specifically as the main mechanism shaping the MZR: as galaxies get more massive, the deeper potential wells  and/or less efficient outflows mean that galaxies retain a higher fraction of the metals produced \citep{dekel86, dalcanton07, dave11, chisholm18}.  Equilibrium models, based on the regulation of star formation by gas inflows and outflows, specifically propose that the gas-phase metallicity is determined by the instantaneous properties of the galaxies, in particular their metal production and cold ISM mass \citep{finlator08, lilly13}. In the model of \citet{forbes14}, it is the recent gas accretion history that sets the present-day gas-phase metallicity of a galaxy. Therefore, while the shape of the MZR encodes information about metal production and feedback processes, its scatter should be related to the natural galaxy-to-galaxy variations at fixed stellar mass in formation/accretion histories and present-day gas contents. 

This expectation from theory has now been verified, in the form of an observed anti-correlation between cold gas mass (both atomic and molecular) and offset from the MZR: at fixed stellar mass, the most gas-rich galaxies are the most metal-poor (and vice-versa) \citep{hughes13,lara-lopez13,bothwell13,jimmy15,bothwell16,brown18,chen22,scholte23}. The same result is also recovered in numerical simulations \citep[e.g.][]{torrey19,delucia20,vanloon21}, where it arises from the self-regulation of the star formation activity via gas accretion and feedback. This important three-way relation between stellar mass, gas mass and metallicity also gives rise to the observed correlation between MZR scatter and star formation rate \citep[e.g.][]{ellison08, lara-lopez10, mannucci10}, via the link between gas contents and star formation activity. This relation is commonly referred to as the "fundamental metallicity relation" (FMR), although both observations and simulations now show that gas mass, rather than SFR, is the more fundamental observable linked to the scatter of the MZR (e.g. \citet{bothwell13, brown18, torrey19, scholte23}, see however \citet{baker23} for the respective roles of SFR and gas mass as the third parameter in the {\it resolved} MZR). 

The aim of this paper is to go one step further and seek observational evidence for the physical drivers of the relation between MZR scatter and gas mass. The most common interpretation invokes variations in gas accretion: at fixed stellar mass, galaxies which have recently accreted a significant amount of (metal-poor) gas will have both lower metallicities and higher gas masses. Conversely, galaxies where gas accretion has been inefficient would have depleted gas reservoirs and higher metallicites. Multiple studies \citep[e.g.][]{yates12,derossi17,trayford19,torrey19,vanloon21} have found through simulations that gas accretion does explain most of the scatter at lower stellar mass, but that at \mstar$ \gtrsim 10^{10.5}$\msun, AGN feedback may be a more important factor.

This is the scenario we are testing in this study. In particular, we use information about the characteristic halo mass of different galaxy sub-samples (using the amplitude of the two-point correlation function as a proxy), since simulations show that accretion rates onto galaxies depend on their large scale environment, including factors such as the mass of their dark matter halo, whether they are the central or a satellite galaxy in this halo, and on the position of the system with respect to the cosmic web \citep{keres05, dekel06, vandevoort11}.  Some studies have so far looked at the impact of large scale structure on the gas-phase metallicites. \citet{donnan22} considered specifically the influence of the position of a galaxy with respect to the cosmic web on its gas phase metallicity, finding that galaxies closest to nodes of the cosmic web are more metal-enriched. Similarly, but with a much smaller sample of galaxies in a specific cosmic filament at $z\sim0.5$, \citet{darvish15} report higher metallicities for galaxies closest to the filament. Other authors have reported an increase in metallicity with the local density of galaxies \citep{mouhcine07, petropoulou12, peng14}. The challenge is in the interpretation of these results, in particular disentangling the relative contributions of the intrinsic and environmental factors, a process that can be eased by drawing upon the results of numerical simulations and models \citep[e.g.][]{peng14, genel16}.

In this work, we provide new insights into the joint relation between stellar mass, gas mass, and metallicity, and the role that gas accretion and halo mass have in shaping it. This is achieved by using a large sample of galaxies with both metallicites and cold gas masses derived from their SDSS optical spectra. The sample selection and physical parameter derivations are explained in Sec. \ref{sec:sample}. The sizeable sample allows us to use galaxy clustering as a proxy for halo mass; these methods are presented in Sec. \ref{sec:methods}, with the results presented and discussed in Sec. \ref{sec:results}. Unless stated otherwise, physical quantities measured and used in this work assume a $\Lambda$CDM cosmology with $\Omega_m = 0.3$ and ${\rm H}_0=70$ \kms~Mpc$^{-1}$, and a Chabrier IMF.

\section{Sample and measurements}
\label{sec:sample}

\subsection{SDSS sample selection and data products}

The sample is selected from SDSS Data Relesase 8 \citep{SDSSDR8}, using emission line fluxes, stellar masses and star formation rates from the MPA-JHU catalogue \citep{brinchmann04,kauffmann03,tremonti04}. From the Main Galaxy Sample \citep[MGS;][]{SDSSMGS} of SDSS, we select galaxies with $0.02 < z < 0.055$ and \mstar$>10^9$~\msun. At $z=0.055$, the SDSS spectroscopic sample is complete for galaxies with \mstar$>10^{9.4}$\msun. We have tested that the results are consistent between the two \mstar\ thresholds for the sample selection, and adopt \mstar$>10^9$~\msun\ since the resulting increase in sample size reduces the uncertainties on the correlation function, especially at small separations (see Sec. \ref{sec:methods}). 
 
As we will be measuring correlation functions, we further restrict the sample to the contiguous sky area delineated by $130^{\circ} < \alpha_{\rm J2000} < 240^{\circ}$ and $0^{\circ} < \delta_{\rm J2000} < 60^{\circ}$. 

Finally, as we wish to study gas-phase metallicity, we remove quiescent galaxies from the sample by rejecting galaxies located more than 0.5~dex below the main-sequence, using the definition of \citet{saintonge22}. In order to measure robust metallicities, we require galaxies to have an \ha\ line flux measured with S$/$N$>15$. Such a high threshold ensures the reliable detection of the four strong lines required to calculate robust metallicities (see Sec. \ref{sec:metallicities}), without introducing metallicity-dependent biases, as would be the case if instead we used the approach of requiring a specific S/N level in each of the four strong lines \citep{cidfernandes10,juneau14}. Given our selection based on distance from the main sequence, the requirement of having S$/$N$>15$ in the \ha\ line only rejects 3\% of the galaxies, and therefore does not bias the sample.

After applying these selection cuts, the sample consists of 36000 galaxies.

Based on their position in the [NII]-BPT diagram \citep{baldwin1981}, we identify star forming galaxies using the \citet{kauffmann03agn} criterion. This "SF sample" includes 31000 galaxies and is used in the analysis in Sec. \ref{sec:MZRscatter} and \ref{sec:clustering}.

To study the possible impact of AGN activity on the scatter of the MZR, in Sec. \ref{sec:AGN} we use as our ``LINER sample" the 2600 galaxies above the \citet{kauffmann03agn} line in the BPT diagram, excluding only Seyfert galaxies (identified as those galaxies with log([NII]/\ha)$>-0.22$ and log([OIII]/\hb)$>0.48$, \citet{ho97}) due to the complications in calculating their metallicities via strong emission lines --- see Sec. \ref{sec:metallicities} for more on this.

\subsection{Gas-phase metallicities}
\label{sec:metallicities}

The observed behaviour of the mass-metallicity relation, in terms of its exact shape, normalisation, scatter and redshift evolution, can be sensitive to the method employed to calculate the metallicity \citep[e.g.][]{kewley08,yates12,curti20}. The main results of this study make use of the metallicity calibration of \citet[][hearafter PP04]{PP04}: 
\begin{equation}
    12+\log({\rm O/H})=8.73-0.32 \times \log \left( \frac{[{\rm O III}] \lambda5007/{\rm H}\beta}{[{\rm N II}] \lambda6584/{\rm H}\alpha} \right)
\label{eq:PP04}
\end{equation}

To test the robustness of the results against metallicity measurement techniques, we also calculate metallicities using the empirical calibration of \citet{D02} which, unlike Eq. \ref{eq:PP04}, only makes use of the [NII]/\ha\ ratio. We also derive metallicities from CLOUDY photoionisation modeling using the implementation of \citet{scholte23}. All results presented in this paper, while using the PP04 calibration, have been verified to be robust when using the other calibrations. Note that while there is relatively good agreement between PP04 and D02 (with differences in 12+$\log$(O/H) of $\sim$0.1 dex, \citep{kewley08}), other metallicity calibrations can have larger discrepancies ($\sim$0.4-0.7 dex \citep[e.g.][]{kewley08,teimoorinia21}. With larger samples of galaxies with robust metallicity measurements from temperature-sensitive indicators, as are now becoming available from the DESI survey for example \citep{zou23}, we will in the future be able to validate the results in this study with yet more varied metallicity indicators.

Caution must be used when estimating gas-phase metallicites in situations where line emission is mostly triggered by shocks or AGN activity rather than star formation. As strong line methods for determining  metallicity are calibrated using H II regions, they may not result in a reliable metallicity when the target is not a star forming region. \citet{sanders17} studied the bias introduced by contamination of emission lines by regions of diffuse interstellar gas (DIG) and flux weighting effects. In the strong-line relations, biases up to $\Delta O3 = +0.3$ dex were found between spectra of mock galaxies and HII regions. 

\citet{kumari19} used HII-DIG region pairs to study the bias that diffuse gas introduces in the calibration of strong line metallicities. Based on these, they then proposed the following correction term to be subtracted from the metallicities of data points beyond the Kauffmann line \citep{kauffmann03agn} on the [NII]-BPT diagram: 
\begin{equation}
\Delta\log(Z)_{O3N2}=-0.127\times O3 - 0.033
\end{equation}
which we have done for our LINER sample. As this correction has not been validated in the regime of Seyfert galaxies, these objects have been excluded from our sample. This method has previously been used successfully by \citet{kumari21} to include LINER-classified galaxies in a study of the FMR.  We note however that there is not a clear consensus on the optimal way to calculate metallicties in LINERs \citep[see e.g.][in contrast to \citet{kumari19}]{oliveira22}, therefore the results obtained in this study for the LINER sample (see Sec. \ref{sec:AGN}) should be considered as encouraging, but requiring confirmation in the future through alternative metallicity measurements, for example using stellar continuum rather than emission lines \citep[e.g.][]{thorne22} or detailed photoionisation modeling \citep[e.g.][]{vidalgarcia22}. 

\begin{figure*}
    \centering
    \subfloat{\includegraphics[width=6cm]{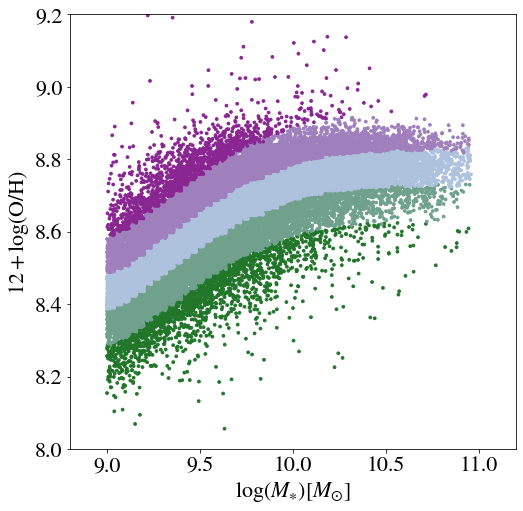}}\hspace*{1.5em}
    \subfloat{\includegraphics[width=11.3cm]{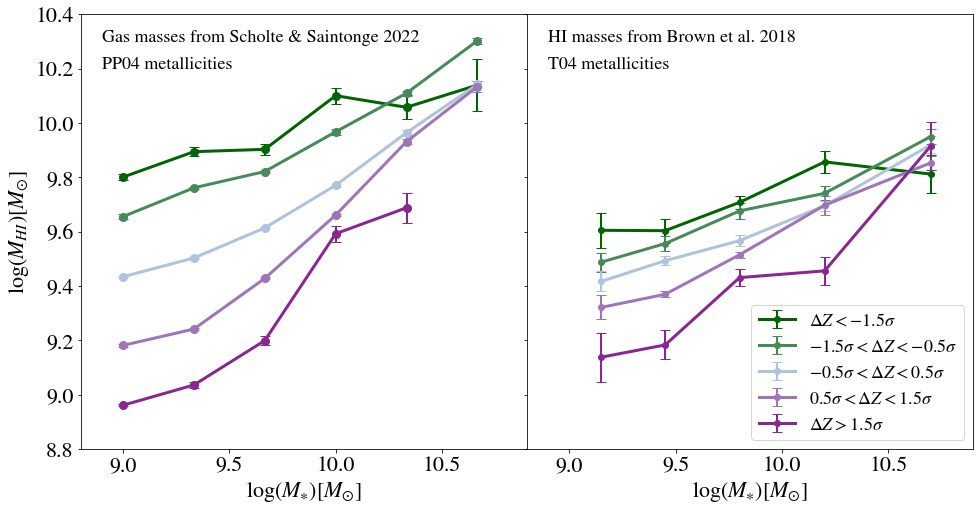}}
    \caption{Relation between mean gas mass and stellar mass (right panels) for star-forming galaxies binned by offset from the MZR (as defined in the left-hand panels). The results are shown for two sets of measurements: the optically-derived gas masses from \citet{scholte23} (see Sec. \ref{sec:gasmasses} for details on this dataset)} and the results obtained by \citet{brown18} from stacking of ALFALFA HI spectra.
    \label{fig:gassmass_comparison}
\end{figure*}

\subsection{Gas masses}
\label{sec:gasmasses}
Despite significant progress in the past decade, cold atomic and molecular gas masses from direct HI and/or CO observations are to this day not available for very large and complete samples such as those that can be assembled from optical samples like SDSS \citep[see review by][]{saintonge22}. Here, we therefore make use of an additional SDSS dataset that includes molecular gas masses (\mh) derived using optical emission lines and photoionisation modeling \citep{scholte23}. The uncertainly on individual cold gas mass measurements obtained through optical emission lines at $\sim0.25$~dex is large compared to other more direct methods, however, the sample size for which gas mass measurements can be derived with this technique is orders of magnitude larger. The principle is that the optical lines encode information about the dust contents of the galaxies, which can be turned into a gas mass estimate \citep{brinchmann13, concas19, yesuf19, piotrowska20}. We refer the reader to \citet{scholte23} for details of the method, but in short, the gas masses are derived from the attenuation of the Hydrogen Balmer lines, in combination with metallicity estimates to constrain the dust-to-gas ratio. Calibration is done against the xCOLD GASS galaxies with CO-based \mh\ measurements \citep{saintonge17}. To allow for comparison with the results of \citet{brown18} in Fig. \ref{fig:gassmass_comparison} and Sec. \ref{sec:MZRscatter}, we infer \mhi\ values from these \mh\ estimates using the relation between \mh$/$\mhi\ and \mstar\ from \citet{catinella18}. It should be noted that the dataset used in Fig. \ref{fig:gassmass_comparison} and Sec. \ref{sec:MZRscatter} is slightly different from the dataset mentioned previously and used throughout the rest of the work. The \citet{scholte23} dataset does not include cuts in $\alpha_{\rm J2000}$ and $\delta_{\rm J2000}$, but does come with a redshift cut of $0.027<z<0.055$ (compared to $0.020<z<0.055$ used in the rest of this study), explaining the very small differences in the MZRs shown in Figures \ref{fig:gassmass_comparison} and \ref{fig:SF_corr}.

\subsection{AGN properties}
\label{sec:BHmasses}
To study the possible effect of AGN feedback on the scatter of the MZR (Sec. \ref{sec:AGN}), we calculate for our LINER sample the quantity:
\begin{equation}
    \lambda_E = \log \Big( \frac{L_{\mathrm{[O III]}}}{\sigma^4} \Big), 
\end{equation}
which serves as proxy for the Eddington ratio, and therefore the rate of accretion onto the supermassive black hole. Note that this equation is only applicable to the inner few kpc of a galaxy. As the SDSS' fibers only covers the innermost 3 kpc for the highest redshift galaxies in our sample, we may use it here. To account for the resolution limit of the SDSS, a velocity dispersion cut of $\sigma>70$ \kms\ was applied.
Following \citet{wild07} and \citet{graves09}, the [O III] luminosity was corrected for dust extinction using the Balmer decrement, and the SDSS fiber stellar velocity dispersion was corrected for aperture effects as:
\begin{equation}
    \sigma=\sigma_{\rm fib}(8 r_{\rm fib}/r_\circ)^{0.04}
\end{equation}
where $ r_\circ = R_{\rm deV}\sqrt{a/b_{\rm deV}} $ is the circularised galaxy radius in arcseconds, and $R_{\rm deV}$ and $a/b_{\rm deV}$ are the radius and axis ratio of the r-band de Vaucouleurs model. For the SDSS spectra, $r_{\rm fib}=1.5^{\prime\prime}$.

\subsection{Clustering amplitude; a proxy for halo mass}
\label{sec:methods} 

As introduced above and discussed in more detail in Sec. \ref{sec:results}, an aim of this study is to link the gas-phase metallicity to the gas contents of galaxies, and specifically with the rate of gas accretion onto the system. We cannot directly measure gas accretion rates with the data available, but simulation shows that it correlates strongly with halo mass \citep[e.g.][]{vandevoort11}. There are on the other hand multiple methods to infer the dark matter halo masses of galaxies, each with its set of advantages and disadvantages: abundance matching \citep{conroy06,behroozi10}, dynamical mass calculations, weak lensing \citep{kaiser95,mandelbaum06,vanuitert16}, and most recently even machine learning \citep{calderon19}. In this study, we use the clustering amplitude of different sub-populations, as measured through the two-point correlation function. Since higher mass halos are more strongly clustered than lower mass halos \citep{sheth99}, at fixed stellar mass, the relative clustering amplitude of different galaxy sub-samples tells us which characteristically reside in the most/least massive halos. In Sec. \ref{sec:clustering}, we also validate our results by considering the abundance matching-derived halo masses and central/satellite classifications from SDSS group catalogs \citep{yang07,lim17}. 

Correlation functions are computed in \verb|Corrfunc| \citep{sinha20}, which uses the \citet{landy1993} estimator
\begin{equation}
    \xi(r_p,\pi)=\frac{DD-2DR+RR}{RR}
\end{equation}
for the redshift-space correlation function. DD, DR and RR are the pair counts of data-data, data-random and random-random. The random catalog was generated to have a random uniform distribution in RA and sin(Dec), with redshift values assigned based on the distribution in the real catalog. 

These are then converted within \verb|Corrfunc| into the projected correlation function
\begin{equation}
    w_p(r_p) = 2 \int_0^\infty \xi(r_p,\pi) d\pi = 2 \sum\limits_{i} \xi(r_p,\pi_i) \Delta \pi_i
\end{equation}
to minimise the effects of redshift space distortion. Here $r_p$ and $\pi_i$ are separations perpendicular and parallel to the line of sight. The upper limit of the summation was set to $\pi_{max}=20 h^{-1}$ Mpc in steps of  $\Delta \pi=2 h^{-1}$ Mpc.

The uncertainty on the projected correlation function is estimated by bootstrapping, 
selecting 80\% of the data at random in each of the $N_b$ draws. The error on the mean of the bootstrapped distribution was then calculated as $\sqrt{N_b/(N_b-1) \times \sigma_{w_p}^2 }$, where $\sigma_{w_p}^2$ is the variance of the projected correlation function across the bootstrap repetitions. 

To compare the strength of the clustering across different galaxy sub-samples, we calculate the relative bias, $b_{rel}(r_{p})$, by comparing the correlation function of a specific sub-sample, $w_{sub}(r_p)$, with the correlation function of the entire sample, $w_{all}(r_p)$: 
\begin{equation}
    b_{rel}(r_{p}) = \sqrt{\frac{w_{sub}(r_p)}{w_{all}(r_p)}}. 
\end{equation}
Following \citet{berti21}, we calculate the relative bias on a "one-halo" scale of 0.1 $h^{-1}$ Mpc < $r_p$ < 1 $h^{-1}$ Mpc and a "two-halo" scale of 1 $h^{-1}$ Mpc < $r_p$ < 10 $h^{-1}$ Mpc. The calculations are made using the default cosmology in \verb|Corrfunc| which has $\Omega_m=0.302$.  By selecting these scales, we can analyse separately effects relating to the relation between satellites and their central galaxies (one-halo scales), and those connected to the clustering of galaxies pairs in different halos (two-halo scales). 

The relative bias is calculated for each bootstrap resampling. We adopt the mean over these repetitions as our measured $b_{rel}$ and calculate the error on this value to be $\sqrt{N_b/(N_b-1) \times \sigma_{b_{rel}}^2 }$, where again $N_b$ is the number of bootstrap repetitions, and $\sigma_{b_{rel}}^2$ is the variance of the relative bias across the $N_b$ repetitions.

\section{Results}
\label{sec:results}

\subsection{Gas mass and the scatter of the mass-metallicity relation: observations and expectations}
\label{sec:MZRscatter}

The mass of the cold ISM of typically star-forming galaxies, both atomic and molecular, has been observed to correlate with the scatter of the MZR; at fixed stellar mass, the more gas-rich a galaxy, the lower its metallicity \citep{bothwell13,bothwell16,brown18}. Analytic models and simulations suggest that this correlation between metallicity and gas mass emerges from the self-regulation of the star formation activity and gas contents of galaxies via accretion of typically metal-poor gas, and outflows of metal-enriched gas \citep{dave12,lilly13,torrey19,delucia20,vanloon21}. The scatter of the MZR therefore contains important information about many components of the baryon cycle; in this study we aim to go one step further and identify which specific processes are most responsible for driving the scatter of the MZR via the gas mass dependence. 

In Figure \ref{fig:gassmass_comparison} we show the median atomic gas mass as a function of stellar mass, in five different ranges of offset from the mass-metallicity relation, \deltamzr. The first important observation is that the results obtained here with optically-derived gas masses (see Sec. \ref{sec:gasmasses}), are qualitatively consistent with the measurements of \citet{brown18} from stacking of \hi\ spectra from the ALFALFA survey \citep{haynes18}. In both sets of results, we observe an anti-correlation between gas mass and metallicity at fixed \mstar, a trend that is strongest at the lowest stellar masses probed ($\sim 10^9$ \msun), gradually weakening as \mstar\ increases. The weakening of the anticorrelation between gas mass and metallicity over the stellar mass range $10^9-10^{10.5}$\msun\ is also seen in the GAEA semi-analytic model \citep{delucia20}. We have checked that the weakening of the anticorrelation in our observations presented in Fig. \ref{fig:gassmass_comparison} is not due to the reduced dynamic range in absolute values of metallicity for a given \deltamzr\ at high \mstar. The results are unchanged after repeating the analysis using \deltamzr\ bins of fixed width instead of fixed quantiles. We therefore show the version of the figure with the fixed quantiles binning; this is consistent with the \citet{brown18} methodology.

Beyond $\sim 10^{10.5}$ \msun, the dependence of \mhi\ on \deltamzr\ disappears, with even indications of a reversal of the trend, with the most metal-poor galaxies at fixed mass now being the most gas-poor. The reversal in the role of gas mass in driving the scatter of the MZR above $10^{10.5}$\msun\ is seen in the EAGLE simulations \citep{vanloon21}, where it can be attributed to AGN feedback \citep{derossi17}. The effect is also reminiscent of the reversal in the role of SFR on the MZR scatter seen by \citet{yates12} and \citet{yates14} in both observations (SDSS) and semi-analytic models (L-Galaxies).

\begin{table*}
    \centering
        \begin{tabular}{lccccccccc}
        \toprule
        \multirow{2}{*}{SF sample} & \multirow{2}{*}{$\Delta Z$} & \multirow{2}{*}{N} & \multirow{2}{*}{$\mathrm{N_{cen}}$} & \multicolumn{3}{c}{12+log(O/H)} & \multicolumn{3}{c}{Bias}\\
        \cmidrule(l{2pt}r{2pt}){5-7} \cmidrule(l{2pt}r{2pt}){8-10}
        & & & & min & mean &  max & one-halo & two-halo & centrals\\
        
        \midrule
        \multirow{6}{*}{$9<\log(M_*/M_\odot)<10$}  &   0.14 & 3931 & 2946 & 8.53 & 8.74 & 9.32 & 1.61(0.10) & 1.28(0.03) & 1.17(0.06)\\
        & 0.07 & 3944 & 2947 & 8.48 & 8.66 & 8.81 & 1.03(0.09) & 1.03(0.04) & 0.98(0.06)\\
        & 0.02 & 3931 & 2946 & 8.44 & 8.62 & 8.77 & 0.98(0.15) & 0.97(0.04) & 0.95(0.07)\\
        & -0.02 & 3941 & 2954 & 8.40 & 8.59 & 8.74 & 1.04(0.15) & 0.95(0.04) & 0.99(0.07)\\
        & -0.06 & 3941 & 2957 & 8.37 & 8.54 & 8.71 & 0.90(0.11) & 0.93(0.04) & 1.00(0.07)\\
        & -0.15 & 3941 & 2954 & 8.20 & 8.46 & 8.67 & 0.83(0.09) & 0.89(0.04) & 0.85(0.08)\\   \hline

        \multirow{4}{*}{$10<\log(M_*/M_\odot)<11$}   & 0.06 & 1828 & 1404 & 8.79 & 8.83 & 9.12 & & 1.03(0.07) & 1.02(0.12)\\
        & 0.02 & 1852 & 1427 & 8.75 & 8.78 & 8.83 &  & 1.00(0.07)  & 0.92(0.12)\\
        & -0.02 & 1839 & 1421 & 8.70 & 8.75 & 8.80 &  & 1.02(0.07) & 1.11(0.13)\\
        & -0.07 & 1862 & 1440 & 8.25 & 8.68 & 8.77 &  & 0.93(0.07) & 0.94(0.13)\\          
        \bottomrule
        \end{tabular}

    \caption{SF sample binned by offset from the MZR, for samples in low and high stellar mass. Given are the metallicity offset, number of galaxies in the bin and the minimum, mean and maximum metallicity of the bin. Relative biases are given on the one-halo and two-halo scale, as well as for a centrals-only dataset.}
    \label{tab:SF_bins}

\end{table*}

\begin{figure*}
    \centering
    \subfloat{\includegraphics[width=7cm]{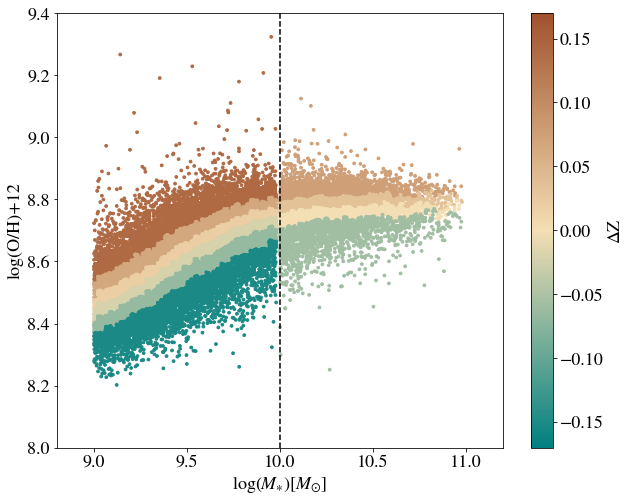}}\hspace*{-.5em}
    \subfloat{\includegraphics[width=11cm]{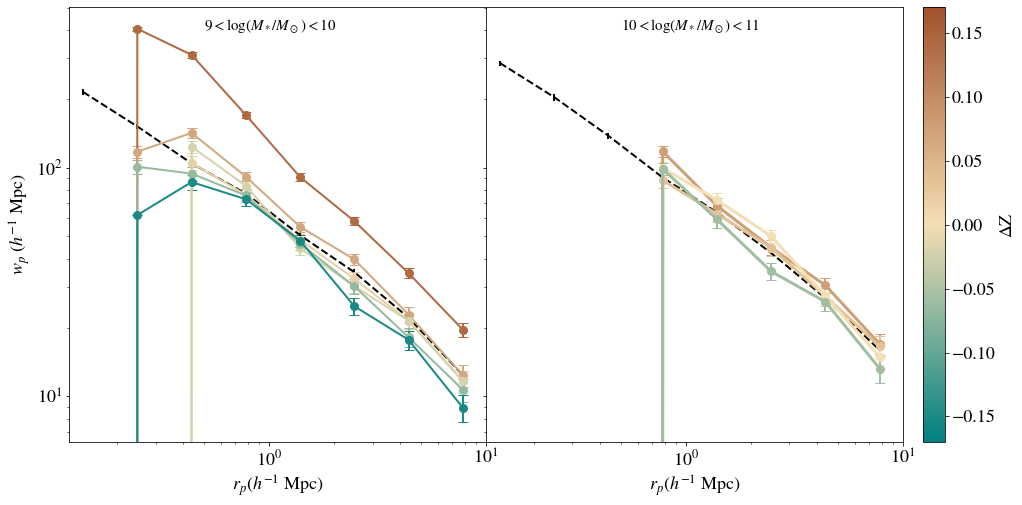}}
    \caption{{\it Left:} Mass-metallicity relation for the "SF sample", showing the binning adopted as a function of \deltamzr, over two different stellar mass ranges. {\it Center and right}: projected correlation functions for each \deltamzr\ subsample, for stellar mass ranges of $9 < \log(M_*) < 10 M_\odot$ and $10 < \log(M_*) < 11 M_\odot$ respectively. The dashed lines represent the correlation function calculated for the full sample within the two stellar mass ranges, with respect to which the relative biases from Table 1 were calculated.}
    \label{fig:SF_corr}
\end{figure*}

The main aim of this study is to look for the physical drivers behind the gas mass-metallicity relation. The results shown in Fig. \ref{fig:gassmass_comparison} not only confirm the existence of this relation, but also strongly suggest that we may be looking for different mechanisms at low and high stellar masses. It is instructive to look at the results of simulations to guide us in this search. 

\subsubsection{Low/intermediate mass galaxies}
In the low- to intermediate-mass regime (i.e. up to \mstar $\sim 10^{10.5}$\msun), there is broad consensus that the scatter in the MZR and its dependence on gas mass are due to fluctuations in accretion rates. Using the IllustrisTNG simulation suite, \citet{torrey19} find that galaxies alternate between being enrichment- and accretion-dominated: periods of increased accretion causing high gas mass and low metallicities are followed by periods of more active star formation that both increase metallicity and decrease gas contents. In this regard, the simulations fully support the simple principles of gas-centric analytic models, where galaxies naturally restore their equilibrium gas mass and metallicity after perturbations to their gas reservoirs \citep{dave12,lilly13,forbes14}. In the EAGLE simulations, \citet{vanloon21} also find that in this \mstar\ range most of the scatter of the MZR comes from systematic variations in gas accretion rates. It is however important to note that these results typically only hold for galaxies whose evolution is dominated by steady gas accretion and secular processes, and not for those held out of equilibrium for an extended period of time by processes such as major mergers and AGN feedback.

In order to know where to look for an observational signature of these gas accretion rate variations, it is necessary to consider their origins and timescales. Generally speaking, the variability in low-metallicity gas accretion rates onto the central star-forming regions can have causes that are either external (accretion of gas onto the galaxy from the large-scale environment) or internal (displacement of gas from extended low density HI envelopes to the denser star-forming disc). In either case, the timescales are expected to be long, on the order of several Gyr, at least at the present time. For example, the timescale expected for the natural up-and-down changes in gas mass and star formation activity under the equilibrium model are long, at $0.4t_H\sim5$~Gyr \citep{lilly13,tacchella16}. Similarly the timescale to transport gas radially from the outer disk to the central region of a galaxy is set by the dynamical friction timescale, which approaches the Hubble time at $z=0$ \citep{tacconi20}, with simulations suggesting a timescale of $\sim5$~Gyr to transport gas radially by 10kpc \citep{okalidis21}. There is also evidence from simulations that MZR offset, and the systematic variations in gas accretion rate, are long-lived (timescales of several Gyr), although we note that \citet{torrey19} report shorter timescale variations (on the order of $\sim1$Gyr) in the gas mass, SFR and metallicity of low mass galaxies.

Our aim here is to look for evidence of this scenario, and investigate whether any changes in gas supply are due to internal or external factors. On the one hand, we expect galaxies in overdense regions of the cosmic web to have less access to fresh gas supply, and therefore to display higher metallicities.  This is seen for example by \citet{donnan22} as an increased metallicity in galaxies closest to nodes in the cosmic web, and in a range of studies that look for the impact of environment on metallicity, which is particularly strong for satellite galaxies \citep[see summary in e.g.][]{maiolino19}. On the other hand, if the link between gas mass and metallicity is caused either by feedback or by internal gas displacement (rather than external accretion), then at fixed stellar mass we might not expect a strong link between gas, metallicity, and halo properties/environment. 

To shed light on the possible roles of internal and external factors, we perform a test in Sec. \ref{sec:clustering}, using two different methods: first we use the clustering amplitude measured from the two-point correlation function as a proxy for the typical halo mass of different galaxy sub-samples, and secondly, by using the central/satellite classification and halo masses from the group catalog of \citet{lim17}.

\subsubsection{High mass galaxies}

In the high mass regime (\mstar $> 10^{10.5}$\msun), both EAGLE and IllustrisTNG simulations find that the scatter of the MZR is no longer driven by systematic variations in gas inflow rate, but instead dominated by the impact of AGN feedback. There should be no strong correlation between clustering and MZR offset, if this is the case.

\subsection{The role of gas accretion: clustering analysis}
\label{sec:clustering}

We use the "SF sample" to calculate the correlation function and compare the clustering amplitude of different galaxy sub-samples as a function of their offset from the MZR. Based on the observed dependence of MZR scatter on gas mass (see Fig. \ref{fig:gassmass_comparison}) and expectations from simulations, we calculate the correlation function over two different stellar mass intervals. To ensure enough statistics in the high-\mstar\ sample, we consider the mass ranges $10^9-10^{10}$\msun\ and $10^{10}-10^{11}$\msun. We have verified that changing the transition point between the two subsamples to $10^{10.5}$\msun\ does not affect the results other than increasing the uncertainty in the high-\mstar\ sample. In each stellar mass interval, the galaxies are divided in equally populated subsamples based on the offset from the MZR (see Fig. \ref{fig:SF_corr} and Table \ref{tab:SF_bins}; 6 bins for the low-\mstar\ subsample, 4 for the high-\mstar\ subsample).

The correlation function for the low- and high-\mstar\ subsamples are shown in the middle and right panels of Fig. \ref{fig:SF_corr}, respectively. The correlation functions show that for the low-\mstar\ subsample, galaxies in the two highest metallicity bins cluster more strongly. This is confirmed by the relative bias for both the one- and two-halo terms.  Conversely, the most metal-poor galaxies are less clustered than the average population, especially on two-halo term scales (Table \ref{tab:SF_bins}). 

For the high-\mstar\ subsample, the correlation functions of the different metallicity sub-samples are more similar. The two-halo relative bias values for the four metallicity bins are consistent with the overall population, within  uncertainties (Table \ref{tab:SF_bins}). If interpreting clustering amplitude as a proxy for the mass of the main halo  \citep{sheth99}, then the conclusion is that the scatter in the MZR at high \mstar\ is not connected to a physical process that depends strongly on halo mass. 

An important consideration is that the galaxy sample used so far includes both central and satellite galaxies. This somewhat complicates the interpretation of the clustering results, because the evolutionary pathways of central and satellite galaxies are different, resulting in different galaxy-halo connections \citep{watson13}. It has also been shown that satellite galaxies have on average higher metallicities than centrals of the same mass \citep{pasquali12, peng14, derossi15}. We therefore need to see whether the difference in clustering observed in Fig. \ref{fig:SF_corr} can be explained by variations in the fraction of satellite galaxies.  To perform this test, we use the SDSS group catalog of \citet{lim17}, which itself is based on the method of \citet{yang07}.  For each galaxy in our "SF sample", the group catalog provides an estimate of the dark matter halo mass, $M_h$, and an indication of whether the galaxy is the most massive in its halo (in which case it is labeled as the central), or else a satellite.  

\begin{figure}
    \centering
    \includegraphics[width=\columnwidth]{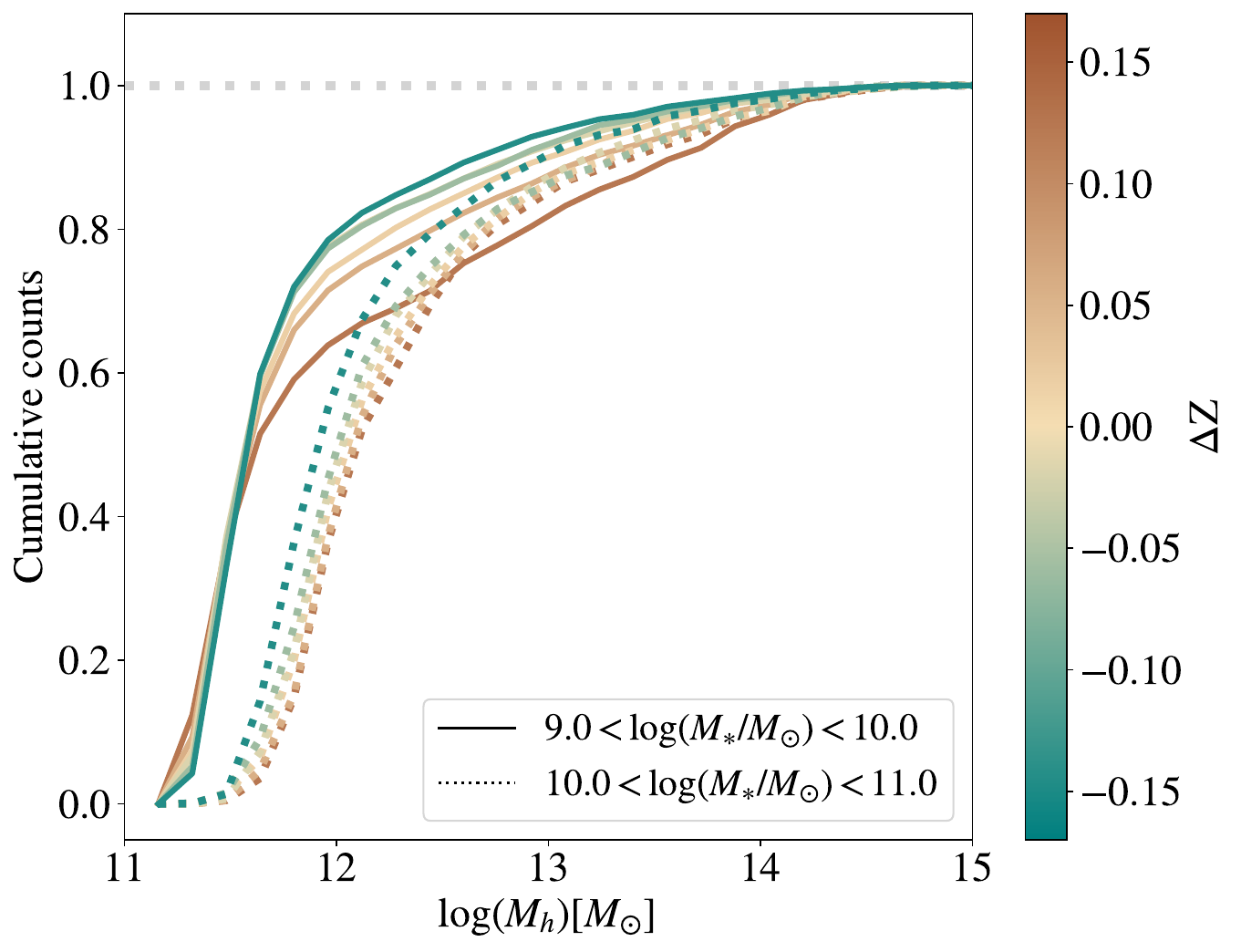}
    \caption{Cumulative distribution of dark matter halo masses (from the \citet{lim17} group catalog) for the low-\mstar\ (solid lines) and high-\mstar\ (dotted lines) subsamples. In each case, the distribution is given for different bins of offset from the MZR, \deltamzr.} 
    \label{fig:Mhalo}
\end{figure}

First, we look in Figure \ref{fig:Mhalo} at the cumulative distribution of halo masses from the \citet{lim17} group catalog for galaxies as a function of \deltamzr. The results are fully consistent with those obtained independently through the clustering analysis in Figure \ref{fig:SF_corr}: at fixed stellar mass, galaxies with the higher/lowest metallicities reside on average in more/less massive halos.

\begin{figure}
    \centering
    \includegraphics[width=\columnwidth]{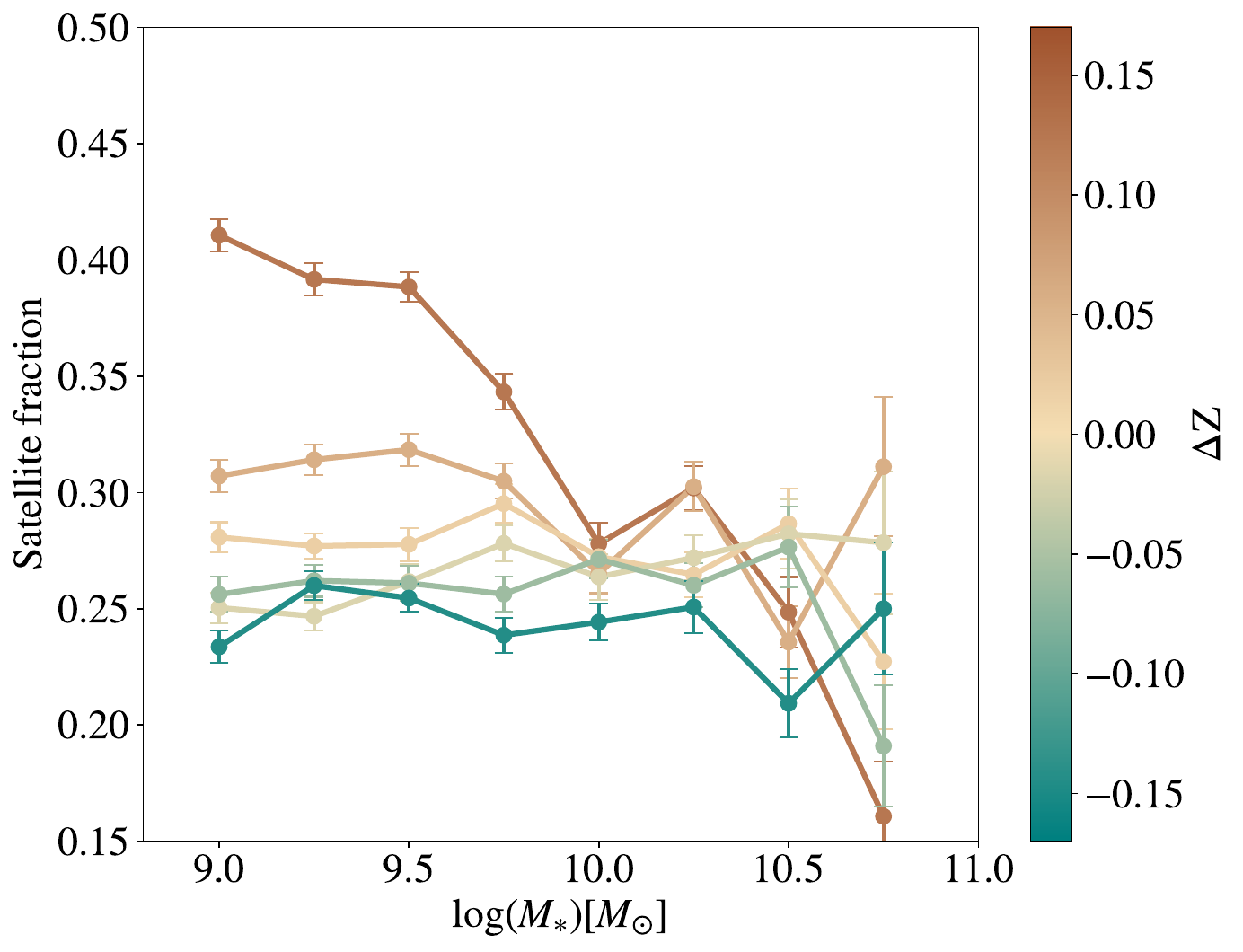}
    \caption{The satellite galaxy fraction as a function of stellar mass, for different sub-samples defined by their offset from the median MZR. The satellite/central galaxy classification is obtained from the \citet{lim17} SDSS group catalog.} 
    \label{fig:fsat}
\end{figure}

Secondly, in Figure \ref{fig:fsat}, we show the fraction of satellite galaxies in each \deltamzr\ subsample. At low stellar masses (\mstar$<10^{10}$\msun), the satellite fraction is significantly enhanced for the most metal-rich subsamples, while at higher stellar masses there is no strong dependency of satellite fraction of \deltamzr. This result is perfectly consistent with the expectations from equilibrium galaxy models: \citet{dave12} make the argument that satellite galaxies are expected to have lower gas content and higher metallicities than centrals at fixed mass, because gas accreted onto the system goes preferentially to the centres of the halos, bypassing the satellites. 

\begin{figure}
    \centering
    \includegraphics[width=\columnwidth]{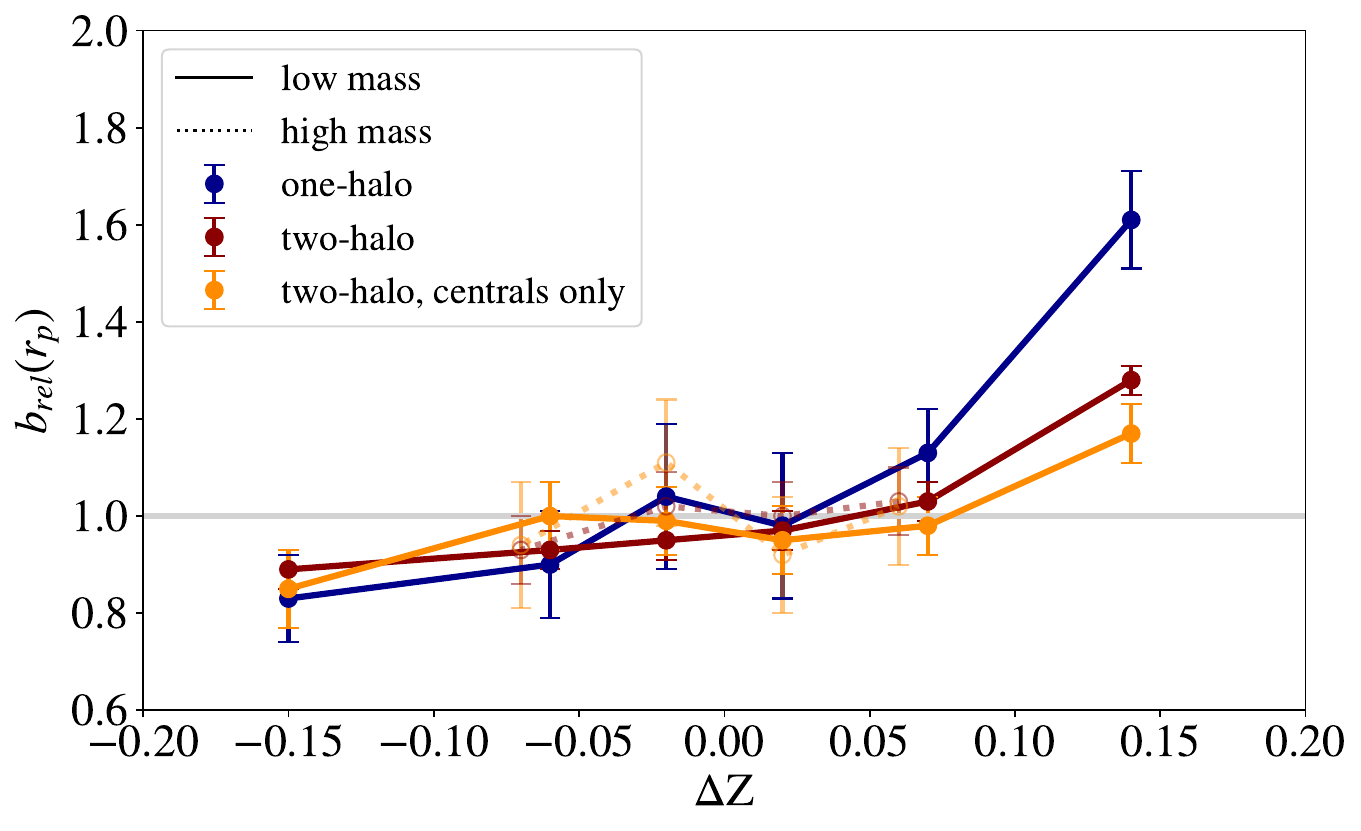}
    \caption{Relative bias on one- and two- halo scales for different \deltamzr\ sub-samples as compared to the entire "SF sample". The results for the low mass subsample ($10^9-10^{10}$\msun) are shown with filled circles and solid lines, the high mass subsample  ($10^{10}-10^{11}$\msun) with open circles and dotted lines. } 
    \label{fig:bias}
\end{figure}

To check whether the results of Fig. \ref{fig:SF_corr} are entirely caused by these variations in the satellite fraction, we calculate the correlation function for central galaxies only, and report the two-halo relative bias values in Table \ref{tab:SF_bins} and Figure \ref{fig:bias}. The amplitude of the bias is reduced, but there is still a statistically meaningful result for the highest and lowest \deltamzr\ bins. 

In summary, as shown in Fig. \ref{fig:bias}, we find that at fixed stellar mass, galaxies with the highest metallicities are significantly more clustered than average. This effect is partly caused by a significantly higher fraction of satellite galaxies amongst these metal-rich and gas-poor galaxies where environment-specific processes such as stripping and strangulation are at play, but the result still holds when considering only central galaxies. Conversely, galaxies with the lowest metallicities at fixed stellar mass are the least clustered. For galaxies within the 1$\sigma$ scatter of the MZR, there is a mild trend between metallicity and clustering amplitude for low mass galaxies (\mstar$<10^{10.0-10.5}$\msun), especially on one-halo scales.

For high mass galaxies (\mstar$>10^{10}$\msun), there is no correlation between \deltamzr\ and clustering. In the next section, we investigate what might instead be responsible for the scatter of the MZR at high masses.

\subsection{AGN influence}
\label{sec:AGN}

\begin{figure}
    \centering
    \includegraphics[width=\columnwidth]{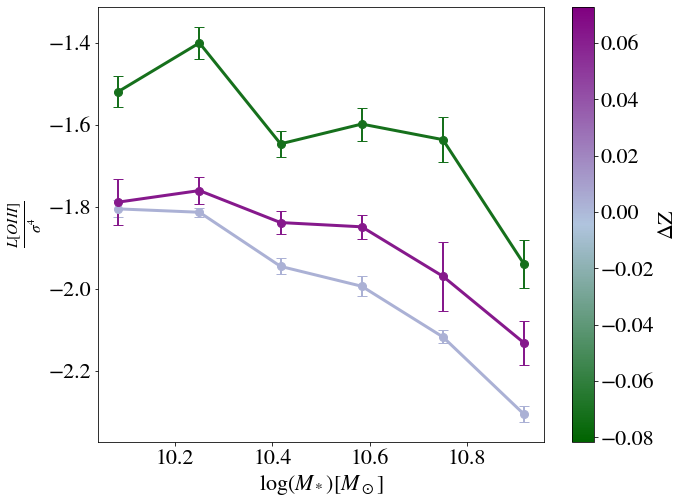}
    \caption{The Eddington ratio for galaxies binned by their offset from the median MZR, as a function of stellar mass. The high/low metallicity galaxies are defined as those that fall above/below one standard deviation from the mean MZR at fixed setellar mass.}
    \label{fig:Le}
\end{figure}

The L-Galaxies, IllustrisTNG and EAGLE simulations \citep{yates12, yates14, derossi17, torrey19, vanloon21} predict that there is a link between the MZR scatter of high stellar mass galaxies ($M_*>10^{10.5}M_\odot$) and AGN feedback. In Sec. \ref{sec:clustering} we used the projected correlation function to show that there is very little correlation between halo mass (and therefore gas accretion, following the line of arguments presented above) and the scatter of the MZR in massive galaxies. To test whether a relation can be found instead with AGN activity, we show in Figure \ref{fig:Le} our proxy for the Eddington ratio, $L_{\rm [OIII]}/\sigma^4$, as a function of \mstar\ for the different \deltamzr\ bins. We consider the "LINER sample" here, which, same as the SF sample from the previous section, does not include galaxies that are more than 0.5 dex below the main-sequence. 

We see a strong correlation at high \mstar\ between AGN activity and \deltamzr: galaxies with the most actively accreting black holes have the lowest metallicity. Under the assumption that AGN with the highest accretion rates also have stronger feedback, there is then some consistency between the simulation predictions and the results shown in Figure \ref{fig:Le}.  \citet{derossi17} interpret the low metallicities of the galaxies with the most active black holes as the result of ejection of preferentially metal-enriched gas. \citet{yates14} provide an explanation for the low metallicities of massive galaxies with large black holes that relates to a slow dilution of the ISM, but that mechanism is less likely to be important in our sample as we have excluded galaxies with low SSFRs where this process dominates. 

We also see in Figure \ref{fig:Le} an intriguing effect: the galaxies with the highest metallicities at fixed mass also have elevated AGN accretion rates (although to a lesser extent), compared to the bulk of the galaxies located within 1$\sigma$ of the MZR. A possible interpretation is that we are seeing the impact of the different timescales regulating AGN activity, star formation and ISM enrichment interacting in an intricate way, but a more thorough analysis would require in depth comparisons with simulations and a larger observational sample to control for additional parameters and bin more finely as a function of MZR offset.

\section{Summary of observations and discussion}

We have used a sample of $\sim36,000$ main-sequence galaxies from SDSS DR8 to study the mass-metallicity relation in the local Universe ($0.02<z<0.055$), and especially the nature of its scatter. The specific aim of this work was to look for the possible physical drivers of this connection between the scatter of the MZR and gas mass. Our main findings are as follows: 
\begin{itemize}
    \item The scatter of the mass-metallicity relation correlates with gas mass below $\log M_* \sim 10^{10.5} M_\odot$; at fixed stellar mass the most gas-rich galaxies are also the most metal-poor, as already found by previous works (e.g. \citet{hughes13,bothwell13}. This link between gas and metallicity is strongest at the lowest stellar masses probed in this work (\mstar$=10^9$\msun) and vanishes at \mstar$\sim10^{10.5}$\msun, in agreement with simulations \citep{delucia20} and results obtained from stacking of HI spectra \citep{brown18}. The novelty here is that our results were obtained from cold gas masses inferred from optical emission lines. While such indirect measures of gas mass have significant uncertainty for individual galaxies, they are robust for statistical studies such as this \citep[see also e.g.][]{piotrowska20,scholte23}.

    \item Given the change in both the shape and scatter of the MZR at \mstar$=10^{10.5}$\msun, we investigate separately low- and high-mass galaxies. At low stellar masses, we find that the most metal-rich galaxies are found on average in more massive dark matter halos (and vice versa), as inferred both by the clustering amplitude in the two-point correlation function and using the halo masses from the \citet{lim17} group catalog. This seems to be in contradiction to the simulations ran by \citet{delucia20}. Looking separately at centrals and satellites shows a contribution of both environmental effects on satellites and the halo mass itself. Since the accretion rate of gas onto galaxies is dependent on halo mass, our results support the picture emerging from simulations and analytic models, which find that the scatter of the MZR and its dependence on gas mass are linked to fluctuations in accretion rates. 

    \item At higher stellar masses, our observations support the prediction from various simulations such as the IllustrisTNG and EAGLE simulations \citep{derossi17,torrey19,vanloon21} that it is no longer gas accretion that drives the scatter of the MZR, but rather AGN feedback. 
    In our sample of LINERs, we indeed find a correlation between the Eddington ratio and MZR offset, with galaxies with the most active black holes having the lowest metallicities. However, the reverse was not shown to be the case for galaxies with the highest metallicities. While we applied a correction to the metallicity measurements to extend their range of applicability into the regime of LINERs, this result should be revisited in the future with alternative metallicity calibrations that are even more robust to the presence of AGN activity. 
\end{itemize}

There is abundant evidence that gas mass is the observable that most closely correlates with the scatter of the MZR, at least at low stellar masses (\mstar$<10^{10.5}$\msun) and in comparison to other parameters such as SFR. This has been observed in the nearby universe using HI- and CO-derived gas masses \citep[e.g.][]{hughes13,bothwell13,bothwell16,brown18}, and now in this work using optically-derived cold gas masses \citep[see also][]{scholte23}. Differences in gas accretion rate are most often invoked to explain this observation, with high (low) gas accretion rate resulting in high (low) gas fraction and decreased (increased) gas-phase metallicity, since the inflowing gas is assumed to be more metal-poor than the ISM. This scenario is supported by simulations and analytic models, but as of yet very little observational evidence has been provided for it, given the considerable challenge of directly measuring gas accretion rates onto galaxies. 

In this study, we showed that at fixed stellar mass, galaxies with the highest (lowest) metallicities are found in dark matter halos that are more (less) massive than average. Above a threshold mass of $M_h \sim 10^{12}$\msun,  simulations show that the rate of gas accretion onto {\em galaxies} (as opposed as into the halo) decreases, the more massive the halo is \citep{vandevoort11}. Under this premise, our observation that the most metal-rich galaxies are found in more massive dark matter halos fully supports the idea that these galaxies have lower accretion rates, leaving them with both lower gas masses and higher metallicities, as observed. Recently, \citet{donnan22} tackled a similar question and report that galaxies closest to nodes in the cosmic web tend to have higher metallicities than average, due to the reduction in low-metallicity gas accretion. They find that their result persists even after controlling for halo mass. The combination of this work and that of \citet{donnan22} therefore suggests that both halo mass itself, and the location of the halo with respect to the cosmic web have an influence on the availability of fresh gas, and therefore on the gas-phase metallicity in the ISM.

When interpreting our results (in particular the correlation between offset from the MZR and halo mass), we have chosen to draw a link between halo mass and gas accretion rate. There are however other physical processes or galaxy properties that correlate with halo mass. For example, at fixed stellar mass, there are systematic variations in halo mass as a function of galaxy morphology and color \citep{mandelbaum06,vanuitert11,taylor20}, as a result of differences in formation histories, gas accretion, and internal secular processes. With larger and more complete spectroscopic samples, such as those now produced by DESI and soon by surveys such as 4MOST-WAVES, it will be possible to repeat the kind of analysis performed here while controlling for some of these additional properties, in order to disentangle the relative impact of external (gas accretion from the cosmic web) and internal processes (radial flows, disc instabilities).  

This study was focused on the local Universe, but even more powerful conclusions will be drawn in the future from tracking changes in the relative roles of environment, gas accretion and feedback in setting the MZR and its scatter as a function of redshift.  There is already evidence that the relation between metallicity and environment disappears or even reverses at $z\gtrsim2$ \citep{kacprzak15, chartab21}. Tracking systematic changes in the nature of the scatter of the MZR with both mass and redshift will be made easier with the next generation of spectroscopic surveys.  Additionally, with JWST allowing us to measure robust metallicities at significantly higher redshifts and in lower mass galaxies than previously possible \citep[e.g.][]{li22,curti22,sanders22}, it will be interesting to push the study of the scatter of the MZR and its dependence on both gas mass and halo mass/environment to these large lookback times where galaxies are expected to be out of equilibrium.

\section*{Acknowledgements}

We thank Rita Tojeiro and Romeel Dav\'e for helpful discussions, and Callum Donnan for looking over the manuscript.  We also thank Toby Brown and Marloes van Loon for making data products available. We are grateful to the UCL Department of Physics and Astronomy for granting this project the Brian Duff Award, and thank the referee for useful comments. 

Funding for SDSS-III has been provided by the Alfred P. Sloan Foundation, the Participating Institutions, the National Science Foundation, and the U.S. Department of Energy Office of Science. The SDSS-III web site is http://www.sdss3.org/. SDSS-III is managed by the Astrophysical Research Consortium for the Participating Institutions of the SDSS-III Collaboration including the University of Arizona, the Brazilian Participation Group, Brookhaven National Laboratory, Carnegie Mellon University, University of Florida, the French Participation Group, the German Participation Group, Harvard University, the Instituto de Astrofisica de Canarias, the Michigan State/Notre Dame/JINA Participation Group, Johns Hopkins University, Lawrence Berkeley National Laboratory, Max Planck Institute for Astrophysics, Max Planck Institute for Extraterrestrial Physics, New Mexico State University, New York University, Ohio State University, Pennsylvania State University, University of Portsmouth, Princeton University, the Spanish Participation Group, University of Tokyo, University of Utah, Vanderbilt University, University of Virginia, University of Washington, and Yale University.

\section*{Data Availability}

The data underlying this article were accessed from the 8th data release of the SDSS survey\footnote{https://www.sdss3.org/dr8/} \citep{SDSSDR8} and the xCOLD GASS survey\footnote{http://www.star.ucl.ac.uk/xCOLDGASS/} \citep{saintonge11,saintonge17}. The derived data generated in this research will be shared on reasonable request to the corresponding author.

\bibliographystyle{mnras}
\bibliography{clustering_refs} 

\bsp	
\label{lastpage}
\end{document}